\documentclass[prd,reprint,superscriptaddress]{revtex4-1}
\usepackage{amsmath}
\usepackage{amssymb}
\usepackage{mathrsfs}
\usepackage{graphicx}
\usepackage[caption=false]{subfig}
\usepackage{enumitem}
\usepackage{color}
\usepackage[percent]{overpic}
\usepackage{bm}
\usepackage{rotating}
\usepackage{hyperref}
\usepackage{cancel}
\usepackage{verbatim}
\usepackage{mathtools}
\usepackage{graphicx}
\usepackage{tensor}
\usepackage{verbatim}
\usepackage{MnSymbol}
\graphicspath{ {images/} }

\usepackage{braket}

\newcommand{\ii}{\mathrm{i}}

\renewcommand{\d}{\mathrm{d}}

\newcommand{\be}{\begin{equation}}
\newcommand{\bel}[1]{\begin{equation}\label{#1}}
\newcommand{\ee}{\end{equation}}

\begin{document}
\title{Purification in Rapid Repeated Interaction Systems}
\author{Daniel Grimmer}
\email{dgrimmer@uwaterloo.ca}
\affiliation{Institute for Quantum Computing, University of Waterloo, Waterloo, ON, N2L 3G1, Canada}
\affiliation{Dept. Physics and Astronomy, University of Waterloo, Waterloo, ON, N2L 3G1, Canada}

\author{Robert B. Mann}
\email{rbmann@uwaterloo.ca}
\affiliation{Dept. Physics and Astronomy, University of Waterloo, Waterloo, ON, N2L 3G1, Canada}
\affiliation{Institute for Quantum Computing, University of Waterloo, Waterloo, ON, N2L 3G1, Canada}
\affiliation{Perimeter Institute for Theoretical Physics, Waterloo, ON, N2L 2Y5, Canada}

\author{Eduardo Mart\'{i}n-Mart\'{i}nez}
\email{emartinmartinez@uwaterloo.ca}
\affiliation{Institute for Quantum Computing, University of Waterloo, Waterloo, ON, N2L 3G1, Canada}
\affiliation{Dept. of Applied Mathematics, University of Waterloo, Waterloo, ON, N2L 3G1, Canada}
\affiliation{Perimeter Institute for Theoretical Physics, Waterloo, ON, N2L 2Y5, Canada}

\begin{abstract}
We investigate the open dynamics of a quantum system when it is rapidly repeatedly updated by a quantum channel. Specifically, we analyze when this dynamics can purify the system. We develop a necessary and sufficient condition for such purification effects to occur and characterize their strength. We thoroughly analyze the specific scenario of a quantum system undergoing rapid unitary interactions with a sequence of ancillary quantum systems. We find that while the purification effects are generally present, in order for these effects to be strong compared to the decoherence effects the interaction Hamiltonian must have a minimum degree of complexity. Specifically, a tensor product interaction $\hat{Q}_\text{S}\otimes\hat{R}_\text{A}$, as well as many common light-matter interactions cannot purify efficiently.
\end{abstract}

\maketitle
\section{Introduction}
The study of the interaction of quantum systems with an unknown environment is relevant to a host of different disciplines ranging from applied physics and engineering to the foundations of quantum theory. For instance, phenomena such as environment-induced decoherence and dephasing hinder our ability to control quantum systems and the flow of quantum information \cite{Nielsen:2000}. Thus it is of capital importance to understand these effects in our efforts to build a quantum computer. Furthermore, such studies are also crucially relevant in our understanding of fundamental topics in quantum theory such as the \textit{measurement problem} \cite{RevModPhys.76.1267}, and more generally, in the context of quantum thermodynamics \cite{kosloff2013quantum}.

When one thinks of the interaction between a quantum system and its environment, the words dephasing (loss of purity) and decoherence come to mind: even if the system and the environment together evolve unitarily, the system's effective dynamics will experience non-unitary evolution. However, not all non-unitary effects decrease purity, so it is thinkable that interaction with an environment can, in principle, also decrease the entropy of the system. Open dynamics can indeed be useful in many different ways. For example, a system could be driven by open dynamics to a fixed point which has some useful property \cite{Layden:2015}, such as enabling entanglement farming \cite{eFarming}.

In this paper we consider the open dynamics that emerges out of the rapid repeated application of a---perhaps stochastic--- completely positive trace preserving (CPTP) map. Within these setups, we will especially focus on the particular CPTP maps generated by the sequential interaction of a system with an ensemble of ancillae. This setup can be thought of as modeling the environment as a sequence of (maybe unknown) constituents which repeatedly couple to the system in rapid succession. In the literature, these scenarios are referred to as \textit{Repeated Interaction Systems} or \textit{Collision Models} \cite{Attal:2006, Attal:2007, Attal:2007b, Vargas:2008, Giovannetti:2012}. These models have been successfully applied to varied phenomena such as, for instance, the study of quantum coherence \cite{Scarani:2002, Ziman:2002, Ziman:2005, Ziman:2005b}, quantum thermodynamics \cite{Bruneau:2006, Bruneau:2008, Bruneau:2008b, Karevski:2009, Bruneau:2014, Bruneau:2014b, Hanson:2015}, the measurement problem (through its close relationship with the quantum Zeno effect, \cite{Layden:2015}) and even decoherence in gravitation \cite{Kafri:2014zsa,Kafri:2015iha, CM} and cosmology
\cite{Altamirano:2016hug}.

Here, we will investigate the particularly interesting possibility that rapid repeated interactions can cause the system's purity to increase rather than introduce decoherence.

Specifically, we will show that that many common types of simple repeated interactions cannot efficiently purify in the rapid interaction regime. We will demonstrate  that the interaction between a system and the constituents of its environment needs to have a minimum degree of complexity in order to cause significant purification. We will identify the necessary and sufficient conditions for a rapid repeated interaction scenario to have significant purification effects on the system. We will also provide particular examples of interactions that can and cannot increase a system's purity under rapid repeated interactions. We will pay special attention to more experimentally relevant setups such as spin J-coupling, the coupling of a qubit to an environment of harmonic oscillators, and we will report particularly surprising results concerning the interaction of the degrees of freedom of electrons in atomic orbitals with relativistic quantum fields, such as the electromagentic field.

This paper is structured as follows: Section \ref{RRIF} reviews the rapid repeated interaction formalism developed in \cite{Layden:2015b,Grimmer2016a}. Section \ref{ECfP} studies when and how strongly rapid repeated interactions can purify. Section \ref{AB} addresses the specific scenario of \textit{ancillary bombardment}. And finally, Section \ref{Examples} presents examples of classes of interactions which can or cannot purify, including the light-matter interaction.


\section{Rapid repeated interactions formalism}\label{RRIF}
In this section we review the results in \cite{Grimmer2016a}, paying close attention to more subtle aspects of their formulation in terms of open dynamics of rapid repeated interactions.

Generally, a rapid repeated interaction scenario consists of a quantum system being frequently updated by a quantum channel. Before we present the formalism for updating by a general quantum channel,  it may be helpful to have a more concrete setup in mind. 

Specifically, a very natural way of thinking about this kind of setup is to consider a quantum system being bombarded by a sequence of ancillary quantum systems, undergoing a brief unitary interaction with each of them. This scenario, which we term \textit{ancillary bombardment}, generates non-trivial open dynamics in the system, as discussed broadly in \cite{Grimmer2016a}. In section \ref{AB} of this paper we analyze this scenario's ability to cause the system state to increase its purity.

As an example, ancillary bombardment could be used to model a system interacting with its environment by assuming that it repeatedly interacts unitarily with individual constituents of the environment. Another example of such a scenario is a laboratory system that is repeatedly bombarded by probes (See \cite{Grimmer2016a} for examples).

With this concrete scenario in mind we can proceed with more formal analysis. A rapid repeated interaction scenario considers a quantum system, labeled S, which evolves (in time steps of duration $\delta t$) by the repeated application of a quantum channel, $\phi(\delta t)$. At each time $T=n \, \delta t$, the  discrete-time evolution of the system's density matrix is given by
\bel{DiscreteMasterEqs}
\rho_\text{S}(n \, \delta t)
\coloneqq\phi(\delta t)^n[\rho_\text{S}(0)],
\ee
for integer $n$. We make the natural assumption that the strength of each individual interaction is finite, so that in the continuous interaction limit, that is as $\delta t\to0$, we have that $\phi(\delta t)\to\openone$ (nothing happens in no time). This is in contrast to approaches where the strength of the interaction is taken to infinity as $\delta t\to0$ \cite{CM,Altamirano:2016knc}  (for an in-depth comparison with previous work see \cite{Grimmer2016a}). Note that since $\phi(\delta t)\to\openone$ as $\delta t\to0$, for small enough $\delta t$, we know $\phi(\delta t)$ is invertible. Additionally, we assume that $\phi(\delta t)$ is differentiable at $\delta t=0$, with derivative $\phi'(0)$ (things happen at a finite rate).

Given such a discrete update map, $\phi(\delta t)$, we can construct a continuous-time interpolation scheme for the dynamics given by \eqref{DiscreteMasterEqs}. Specifically, we find a unique interpolation scheme by making the following three assumptions for the continuous-time evolution: 
\begin{enumerate}
\item The evolution is Markovian, such that,
\bel{IntegratedMasterEqs}
\rho_\text{S}(t)
\coloneqq\exp(t \, \mathcal{L}_{\delta t})[\rho_\text{S}(0)],
\ee
or equivalently,
\bel{MarkMasterEqs}
\frac{\d}{\d t}\rho_\text{S}(t)=\mathcal{L}_{\delta t}[\rho_\text{S}(t)],
\ee
where $\mathcal{L}_{\delta t}$ (the effective time-independent Liouvillian) is some superoperator which generates time translations for the system.

\item The evolution exactly matches the discrete dynamics \eqref{DiscreteMasterEqs} at the end of every time step. Using \eqref{IntegratedMasterEqs} this means,
\be
\exp(n \, \delta t \, \mathcal{L}_{\delta t})
=\phi(\delta t)^n
\ee
or equivalently,
\bel{Matching Condition}
\exp(\delta t \, \mathcal{L}_{\delta t})
=\phi(\delta t).
\ee

\item The evolution's effective Liouvillian, $\mathcal{L}_{\delta t}$, is well defined in the continuous interaction limit, that is as $\delta t\to0$.
\end{enumerate}

These three conditions uniquely specify the interpolation scheme that is generated by
\bel{Ldef}
\mathcal{L}_{\delta t}
\coloneqq\frac{1}{\delta t}\text{log}(\phi(\delta t)),
\ee
where we have taken the logarithm's principal branch cut, that is the one with $\text{log}(\openone)=0$. Note that our assumption that $\phi(\delta t)\to\openone$ as $\delta t\to0$ guarantees that $\phi(\delta t)$ will be nonsingular in the short time regime, and hence will have a well defined logarithm.

The first condition guarantees that the interpolation scheme is generated by some effective time-independent Liouvillian, and the second condition forces this Liouvillian to have the form \eqref{Ldef}. The third condition resolves the ambiguity of the logarithm's branch cut by forcing $\text{log}(\openone)=0$, which is necessary to make $\mathcal{L}_{\delta t}$ well defined as $\delta t\to0$. Moreover, this branch cut allows us to calculate $\mathcal{L}_{\delta t}$ as $\delta t\to0$ (using L'H\^opital's rule) to be
\begin{align}
\mathcal{L}_0
\coloneqq
\lim_{\delta t\to0}\mathcal{L}_{\delta t}
&=\frac{\d}{\d \, \delta t}
\bigg\vert_{\delta t=0}
\text{log}(\phi(\delta t))\\
\nonumber
&=\phi^{-1}(0) \, \phi'(0)\\
\nonumber
&=\phi'(0).
\end{align}
Thus, in the continuum limit, evolution is generated by the derivative of the update map. This result was first explicated in \cite{Layden:2015b}.

Taking all this into account, we can faithfully describe the discrete-time evolution, \eqref{DiscreteMasterEqs}, of a quantum system using the continuous-time interpolation scheme \eqref{MarkMasterEqs}, generated by \eqref{Ldef}.

If in addition to the minimal regularity assumed above (that is, $\phi(\delta t)\to\openone$ as $\delta t\to 0$ and $\phi'(0)$ exists), we also have that $\phi(\delta t)$ is analytic at $\delta t=0$,  we can then form the series expansion
\bel{PhiSeries}
\phi(\delta t)
=\openone
+\delta t \, \phi_1
+\delta t^2 \, \phi_2
+\delta t^3 \, \phi_3
+\dots 
\ee
and from this
\bel{LdtSeries}
\mathcal{L}_{\delta t}
=\mathcal{L}_0
+\delta t \, \mathcal{L}_1
+\delta t^2 \, \mathcal{L}_2
+\delta t^3 \, \mathcal{L}_3
+\dots \, .
\ee
As shown in \cite{Grimmer2016a}, the first few superoperator coefficients are given recursively by
\begin{align}
\label{L0def}
\mathcal{L}_0
\coloneqq\phi_1&,\\
\label{L1def}
\mathcal{L}_1
\coloneqq\phi_2
&-\frac{1}{2}\mathcal{L}_0{}^2,\\
\label{L2def}
\mathcal{L}_2
\coloneqq\phi_{3}
&-\frac{1}{2}(\mathcal{L}_0\mathcal{L}_1+\mathcal{L}_1\mathcal{L}_0)
-\frac{1}{6}\mathcal{L}_0{}^3\\
\label{L3def}
\mathcal{L}_3
\coloneqq\phi_{4}
&-\frac{1}{2}(\mathcal{L}_0\mathcal{L}_2+\mathcal{L}_2\mathcal{L}_0)\\
&\nonumber
-\frac{1}{6}(\mathcal{L}_0{}^2\mathcal{L}_1+\mathcal{L}_0\mathcal{L}_1\mathcal{L}_0+\mathcal{L}_1\mathcal{L}_0{}^2)
-\frac{1}{24}\mathcal{L}_0{}^4.
\end{align}
with the higher order terms following a similar pattern. 

From the series \eqref{LdtSeries}, the master equation for the interpolation scheme \eqref{MarkMasterEqs} becomes,
\bel{MasterEqsSeries}
\frac{\d}{\d t}\rho_\text{S}(t)
=\mathcal{L}_0[\rho_\text{S}(t)]
+\delta t \, \mathcal{L}_1[\rho_\text{S}(t)]
+\delta t^2\mathcal{L}_2[\rho_\text{S}(t)]
+\dots \, .
\ee

Given such an update map $\phi(\delta t)$ we can compute these coefficient maps and analyze their effects in the system dynamics. For instance, in the case of a ancillary bombardment defined above, $\mathcal{L}_0$ generates unitary dynamics \cite{Layden:2015b, Grimmer2016a}. Thus within this model any decoherence effects require finite interaction times. In \cite{Grimmer2016a} it was shown that decoherence effects generically appear in $\mathcal{L}_1$, that is at first order in $\delta t$.

In the following sections we will analyze under what conditions rapid repeated interactions can increase the purity of a system, rather than just introducing decoherence.

\section{Purification Conditions}\label{ECfP}

In this section, we find a necessary and sufficient condition for when the discrete dynamics given by \eqref{DiscreteMasterEqs} can cause purification of a finite dimensional system. By this we mean that there exists some system state, $\rho_\text{S}$, whose purity, $\mathcal{P}(\rho_\text{S})=\text{Tr}(\rho_\text{S}{}^2)$, increases under these dynamics.

In section \ref{RRIF} we converted the discrete-time dynamics \eqref{DiscreteMasterEqs} into the continuous-time Markovian dynamics \eqref{MarkMasterEqs}, generated by the effective Liouvillian \eqref{Ldef}. We will now discuss the exact conditions for such an interpolation scheme to cause purification, show that this interpolation scheme purifies if and only if the discrete dynamics does too, and finally characterize the strength of such purification effects.

\subsection{Markovian Purification}\label{MarkPure}

For  finite $d$-dimensional systems the dynamics generated by a Liouvillian, $\mathcal{L}$, can cause purification if and only if the dynamics which it generates is not unital, that is,
\bel{PurificationCondition}
\mathcal{L}[I]\neq0,
\ee
where $I$ is the \mbox{$d$-dimensional} identity matrix \cite{Lidar2006}. Recalling that the maximally mixed state is given by \mbox{$\rho=I/d$}, we can restate this as: Markovian dynamics can purify if and only if it \textit{moves} the maximally mixed state. Throughout this paper we will refer to $I$ and the maximally mixed state synonymously.

The condition \eqref{PurificationCondition} is clearly sufficient for the dynamics to cause purification since if the maximally mixed state is moved by the dynamics, its purity must increase. This follows from the maximally mixed state being the unique minimum purity state. 

Note, however, that this is not true for infinite dimensional systems. The question of purification of infinite dimensional systems under Markovian dynamics has been analyzed in depth \cite{Lidar2006}, with the result that $\mathcal{L}$ not being unital is still necessary for causing purification, but is no longer sufficient.

The necessity of \eqref{PurificationCondition} to cause purification follows from the claim \cite{Lidar2006}
\bel{PurityRateBound}
\frac{\d}{\d t}\mathcal{P}(\rho)
=\frac{\d}{\d t}\text{Tr} \, (\rho{}^2)
\leq\text{Tr} \, \big(\mathcal{L}[I] \, \rho^2\big)
\ee
whose proof we reproduce with our notation in Appendix \ref{Necessity}.

\subsection{Interpolation Faithfully Captures Purification Effects}\label{Faith}

In the following, we prove that, in the rapid interaction regime, the interpolation scheme \eqref{MarkMasterEqs} faithfully captures the presence of purification effects in the discrete dynamics \eqref{DiscreteMasterEqs}.

First we argue that any purification effects in the discrete dynamics is captured by the interpolation scheme. Suppose that applying the update map, $\phi(\delta t)$, increases the purity of some state. By construction, applying the interpolation scheme for a duration $\delta t$ to this state has to yield the same result. Because the interpolation scheme is smooth, at some point in this duration it must have increased some state's purity. 

Next, we consider the possibility that the interpolation scheme could indicate that there is purification when none is present in the discrete dynamics. Suppose that the interpolation scheme \eqref{MarkMasterEqs} instantaneously purifies some state. Then by the purification condition discussed in section \ref{MarkPure}, we must have $\mathcal{L}_{\delta t}[I]\neq0$. From the matching condition \eqref{Matching Condition}, this implies that 
\begin{align}
\phi(\delta t)[I]
&=\exp(\delta t \, \mathcal{L}_{\delta t})[I]\\
&\nonumber
=I+\delta t \, \mathcal{L}_{\delta t}[I]
+(\delta t^2/2) \, \mathcal{L}_{\delta t}[\mathcal{L}_{\delta t}[I]]
+\dots \, .
\end{align}
and we therefore conclude $\phi[I]\neq I$.   For interaction times small enough, we can neglect the $\mathcal{O}(\delta t^2)$ terms. 

Thus the discrete dynamics do in fact purify, the discrete update map moving (and hence purifying) the maximally mixed state.  Note that this argument relies on the maximally mixed state being the unique minimum purity state and so does not work in infinite dimensions.

From these arguments, we conclude that in the rapid repeated interaction regime, the discrete dynamics generated by $\phi(\delta t)$ can purify if and only if the continuous dynamics generated by $\mathcal{L}_{\delta t}$ can.  It then follows from \eqref{Ldef}  that repeated applications of $\phi(\delta t)$ can purify if and only  \mbox{$\phi(\delta t)[I]\neq I$} (or in other words $\phi(\delta t)$ is not unital).

\subsection{Purification Strength}\label{PStrength}

Now that we have identified a necessary and sufficient condition for the dynamics generated by repeated applications of $\phi(\delta t)$ to purify, we will quantify the strength of this purification. Assuming that the quantum channel, $\phi(\delta t)$ is analytic at $\delta t=0$, we can make use of the series expansion \eqref{LdtSeries} to quantify this strength by noting at what order in $\delta t$ the maximally mixed state is moved by the effective Liouvillian, $\mathcal{L}_{\delta t}$. We say that the dynamics purifies at order $m$ if,
\bel{PurificationOrder}
\mathcal{L}_{\delta t}[I]
=\mathcal{O}(\delta t^m) 
\ee
where $\text{ord}(\phi)\coloneqq m$ is the \textit{purification order} of the dynamics. The smaller $\text{ord}(\phi)$ is the stronger the purification effects. Recall that in infinite dimensions, dynamics being non-unital is not sufficient for purification effects. Therefore this measure of purification strength only makes sense for finite dimensional systems.

This notion of purification strength can be translated to the discrete updater $\phi(\delta t)$ by using the recursive structure of the coefficient maps in \eqref{LdtSeries}. Concretely,
\bel{Offset}
\mathcal{L}_{\delta t}[I]
=\mathcal{O}(\delta t^m)
\Leftrightarrow
\phi(\delta t)[I]
=I+\mathcal{O}(\delta t^{m+1}),
\ee
such that the orders of the non-unital effects are offset by one between the discrete update map and the interpolation scheme.

In order to make use of purification effects experimentally---for example in an algorithmic cooling setup \cite{Rodriguez2016a},--- we would like to manufacture interactions which can purify at the lowest possible order. One may wonder if this is possible by combining different interaction maps to engineer a new map with a lower purification order.  

However, two simple ways of combining maps together, namely concatenation of different maps or applying maps out of a statistical ensemble (taking convex combinations), cannot lower the resultant purification order below those of the original maps. 

Specifically, if we take $\phi(\delta t)$ to be a concatenation of a finite number of maps as,
\bel{Concatenation}
\phi(\delta t)=
\chi^{(1)}(\delta t) \, \chi^{(2)}(\delta t) \, \dots \, \chi^{(N)}(\delta t),
\ee
then 
\be
\text{ord}(\phi)\geq\text{min}\{\text{ord}(\chi^{(n)})\},
\ee
such that $\phi$'s strength is bounded by the strongest $\chi$. Additionally, taking $\phi(\delta t)$ to be a convex combination of maps as,
\bel{ConvexCombo}
\phi(\delta t)=
\sum_k p_k \, \psi^{(k)}(\delta t),
\ee
with $\sum_k p_k =1$ we find
\be
\text{ord}(\phi)\geq\text{min}\{\text{ord}(\psi^{(k)})\},
\ee
such that $\phi$'s strength is bounded by the strongest $\psi$. We prove these claims in Appendix \ref{CaCC}.

\section{Ancillary Bombardment}\label{AB}
We now apply the characterization of purification effects developed in the previous section to a specific physically motivated class of update maps given by, 
\bel{TimelessPhi}
\phi(\delta t)[\rho_\text{S}]
=\text{Tr}_\text{A}\Big(\exp(-\ii \, \delta t \, \textrm{ad}_{\hat{H}}/\hbar)(\rho_\text{S}\otimes \rho_\text{A})\Big)
\ee 
where $\textrm{ad}_{\hat{H}}(A) = [\hat{H}, A]$ for any operator $A$. Physically, this map describes the system, S, first engaging with an ancilla, A, which is in the state $\rho_\text{A}$, then interacting for a time $\delta t$ under the joint Hamiltonian
\bel{TimelessHam}
\hat{H}=\hat{H}_\text{S}\otimes\openone
+\openone\otimes \hat{H}_\text{A}
+\hat{H}_\text{SA},
\ee
and finally decoupling from ancilla, which is discarded.

This update map could be used to model a wide variety of scenarios. For example, it could model each discrete step of the dynamics of a system repeatedly interacting with the constituents of its environment, or an atom being bombarded with light/other atoms in a laboratory setting (both examples of ancillary bombardment).

Note that the necessary and sufficient condition to cause purification, \eqref{PurificationCondition}, which was discussed in section \ref{ECfP}, requires that S be finite dimensional. However, there is no such restriction on the ancillary systems, A, to which S couples.

This update map is sufficiently well behaved in the rapid interaction limit --- recall that we require $\phi(\delta t)\to\openone$ as $\delta t\to0$ and that $\phi'(0)$ exists --- and so we can construct the unique Markovian interpolation scheme as prescribed in section \ref{RRIF}. Moreover, since the update map is analytic around $\delta t=0$, we can expand it in powers of $\delta t$ as in \eqref{PhiSeries}:
\begin{align}
\label{TimelessPhi1}
&\phi_1[\rho_\text{S}]
=\frac{-\ii}{\hbar} \, 
\text{Tr}_\text{A}\Big(
[\hat{H},\rho_\text{S}\otimes \rho_\text{A}]\Big)\\
\label{TimelessPhi2}
&\phi_2[\rho_\text{S}]
=\frac{1}{2!}\Big(\frac{-\ii}{\hbar}\Big)^2
\text{Tr}_\text{A}\Big(
[\hat{H},[\hat{H},\rho_\text{S}\otimes \rho_\text{A}]]\Big)\\
\label{TimelessPhi3}
&\phi_3[\rho_\text{S}]
=\frac{1}{3!}\Big(\frac{-\ii}{\hbar}\Big)^3
\text{Tr}_\text{A}\Big(
[\hat{H},[\hat{H},[\hat{H},\rho_\text{S}\otimes \rho_\text{A}]]]\Big)\\
\label{TimelessPhi4}
&\phi_4[\rho_\text{S}]
=\frac{1}{4!}\Big(\frac{-\ii}{\hbar}\Big)^4
\text{Tr}_\text{A}\Big(
[\hat{H},[\hat{H},[\hat{H},[\hat{H},\rho_\text{S}\otimes \rho_\text{A}]]]]\Big)
\end{align}
and so on. We can thus expand the effective Liouvillian as in \eqref{LdtSeries}.

In \cite{Layden:2015b,Grimmer2016a}, a general family of update maps including \eqref{TimelessPhi} were analyzed at zeroth and first order using the rapid repeated interaction formalism discussed in section \ref{RRIF}. The full generality of the interactions considered in \cite{Grimmer2016a} includes allowing time dependence in the interaction Hamiltonian as well as taking an arbitrary convex combination of multiple interaction types, with different types of ancilla and with different couplings.

Remarkably, in \cite{Layden:2015b}, it was found that $\mathcal{L}_0$ generates unitary evolution. For the dynamics generated by \eqref{TimelessPhi} we have,
\begin{align}\label{TimelessL0explicit}
\mathcal{L}_0[\rho_\text{S}]
&=\frac{-\ii}{\hbar} \, 
[\hat{H}_\text{eff},\rho_\text{S}]
\end{align}
where the effective Hamiltonian $\hat{H}_\text{eff}$ is given by
\bel{H0effdef}
\hat{H}_\text{eff}
\coloneqq
\hat{H}_\text{S}+\hat{H}^{(0)},
\ee 
that is, the system's free Hamiltonian plus a new term, $\hat{H}^{(0)}$, which comes from the repeated interactions. This new contribution to the dynamics is given by,
\bel{H0def}
\hat{H}^{(0)}
\coloneqq
\text{Tr}_\text{A}(\hat{H}_\text{SA}\rho_\text{A}).
\ee
Note that since the leading order dynamics is unitary, it cannot affect the purity of the system. Thus any decoherence effects must arise at subleading order in the dynamics. The leading possible order for purification effects is thus first order.

The first order dynamics, $\mathcal{L}_1$, was analyzed in full detail in \cite{Grimmer2016a}, and was generally seen to give rise to dephasing effects. For the dynamics generated by \eqref{TimelessPhi}, $\mathcal{L}_1$ is given by 
\begin{align}\label{L1DefAB}
\mathcal{L}_1[\rho_\text{S}]
&=\frac{-\ii}{\hbar}
\big[\hat{H}^{(1)},\rho_\text{S}\big]
-\frac{1}{2}\Big(\frac{-\ii}{\hbar}\Big)^2 \, \big[\hat{H}^{(0)},[\hat{H}^{(0)},\rho_\text{S}]\big]\\
&\nonumber
+\frac{1}{2}\Big(\frac{-\ii}{\hbar}\Big)^2 \,
\text{Tr}_\text{A}\Big(
\big[\hat{H}_\text{SA},
[\hat{H}_\text{SA},
\rho_\text{S}\otimes\rho_\text{A}]\big]\Big)
\end{align}
where,
\bel{H1def}
H^{(1)}=\frac{-\ii}{2\hbar}\text{Tr}_\text{A}\big(
\hat{H}_\text{SA} \, [\hat{H}_\text{A},\rho_\text{A}]\big).
\ee
The first order dynamics, $\mathcal{L}_1[\rho_\text{S}]$, consists of two different contributions. One is a new unitary contribution to the dynamics, $\hat{H}^{(1)}$, which (after examination of \eqref{H0def} and \eqref{H1def}) can be understood as correction to $\hat{H}^{(0)}$ accounting for the ancilla evolving under its free Hamiltonian during the interaction. Secondly, there are two other terms that are not unitary and will, in general, affect the purity of the system.

Since \eqref{L1DefAB} generically introduces dephasing effects at order $\delta t$, for any purification effects to have a comparable impact on the dynamics they must also appear at first order. That is\begin{align}\label{L1oI}
\mathcal{L}_1[I]
=\frac{1}{2}\Big(\frac{-\ii}{\hbar}\Big)^2 \,
\text{Tr}_\text{A}\Big(
\big[\hat{H}_\text{SA},
[\hat{H}_\text{SA},
I\otimes\rho_\text{A}]\big]\Big)\neq0 
\end{align}
 or in other words $\mathcal{L}_1$ must already be able to move the maximally mixed state.
Note that this just depends on the interaction Hamiltonian and the state of the ancilla, and not on either of the free Hamiltonians.

In the following subsections we investigate the algebraic conditions that an interaction Hamiltonian needs in order to purify at leading possible order, that is to satisfy \eqref{L1oI}.

\subsection{Tensor Product Interaction}\label{TPI}
We begin by analyzing the simplest model for an interaction Hamiltonian, namely the tensor product of scalar observables. The joint Hamiltonian under this type of coupling is,
\bel{HsaJsJa}
\hat{H}=\hat{H}_\text{S}\otimes\openone
+\openone\otimes \hat{H}_\text{A}
+\hat{Q}_\text{S}\otimes \hat{R}_\text{A},
\ee
where $\hat{Q}_\text{S}$ and $\hat{R}_\text{A}$ are observables of the system and ancilla respectively. This type of interaction is a very common interaction model considered throughout the literature of rapid repeated interaction \cite{CM,Altamirano:2016knc}.
  
In Appendix \ref{TPCalc}, we show that the effect of the first order dynamics   on the maximally mixed state vanishes, $\mathcal{L}_1[I]=0$. Thus rapid repeated interaction under the Hamiltonian \eqref{HsaJsJa} cannot purify at leading order in decoherence effects. In fact we also show that the second order effects vanish, $\mathcal{L}_2[I]=0$. Continuing on, we find the leading order purification effect is given by
\be
\mathcal{L}_3[I]
=\frac{1}{12\hbar^4}
[\hat{Q}_\text{S},[\hat{H}_\text{S},\hat{Q}_\text{S}]] \, 
\text{Tr}_\text{A}\Big([\hat{R}_\text{A},[\hat{H}_\text{A},\hat{R}_\text{A}]]\rho_\text{A}\Big).    
\ee
Note that if the ancillae are infinite dimensional then the above calculations require that any relevant permutations of $\hat{R}_\text{A}$, $\hat{H}_\text{A}$, and $\rho_\text{A}$ are trace class.

Thus a tensor product interaction will in general only be able to purify at third order, that is two orders lower than the leading order decoherence effects. The conclusion of this analysis is that, perhaps unintuitively, a tensor product interaction of the kind $\hat{H}_\text{SA}=\hat{Q}_\text{S}\otimes \hat{R}_\text{A}$ will in general strictly decrease purity at leading order in decoherence effects. This analysis allows us to conclude that any rapid repeated tensor product interaction model cannot capture phenomena involving an entropy decrease in S, such as cooling.

\subsection{Interaction via non-product Hamiltonians}\label{NPIH}
After having established that tensor product Hamiltonians cannot purify at leading order in dechoerence effects, we now investigate whether it is possible to do so through rapid repeated interaction under a Hamiltonian that is the sum of two scalar couplings, i.e.,
\bel{TwoHam}
\hat{H}_\text{SA}
=\hat{Q}_\text{S}\otimes \hat{R}_\text{A}
+\hat{S}_\text{S}\otimes \hat{T}_\text{A}.
\ee
In Appendix \ref{TPCalc}, we show that the effect of $\mathcal{L}_1$ on the maximally mixed state, \eqref{L1oI}, is,
\begin{align}
\mathcal{L}_1[I]
&=\frac{1}{(\ii\hbar)^2}
[\hat{Q}_\text{S},\hat{S}_\text{S}] \ 
\text{Tr}_\text{A}\Big([\hat{R}_\text{A},\hat{T}_\text{A}]\rho_\text{A}\Big).
\end{align}
Again, if the ancillae are infinite dimensional, then the above calculation requires that all relevant permutations of $\hat{R}_\text{A}$, $\hat{T}_\text{A}$, and $\rho_\text{A}$ are trace class.

In contrast to the simple interaction \eqref{HsaJsJa}, the interaction Hamiltonian \eqref{TwoHam} can purify at leading order in decoherence effects. Specifically it will purify if and only if the two system observables ($\hat{Q}_\text{S}$ and $\hat{S}_\text{S}$) do not commute, and the two ancilla observables ($\hat{R}_\text{A}$ and $\hat{T}_\text{A}$) do not commute on average with respect to the initial state of the ancilla, $\rho_\text{A}$.

From this we can move to the most general case, by noting that any interaction Hamiltonian, $H_\text{SA}$, can be decomposed as a sum of tensor products  
\bel{HsaDecomp}
\hat{H}_\text{SA}=\sum_j \hat{Q}_{\text{S},j}\otimes \hat{R}_{\text{A},j}
\ee
In Appendix \ref{TPCalc} we show that, for the general case of \eqref{HsaDecomp}, the effect of $\mathcal{L}_1$ on the maximally mixed state, \eqref{L1oI}, is
\bel{SumProduct}
\mathcal{L}_1[I]
=\frac{1}{2}\Big(\frac{-\ii}{\hbar}\Big)^2
\sum_{i,j}[\hat{Q}_{\text{S},i},\hat{Q}_{\text{S},j}] \ 
\text{Tr}_\text{A}\Big([\hat{R}_{\text{A},i},\hat{R}_{\text{A},j}]\rho_\text{A}\Big)
\ee
and as before, if the ancillae are infinite dimensional, then the above calculation requires that all relevant permutations of $\hat{R}_{\text{S},i}$, $\hat{R}_{\text{S},j}$, and $\rho_\text{A}$ are trace class.

Thus the condition that an interaction Hamilton to be able to purify at leading possible order is,
\bel{GenCond}
\sum_{i,j}
\text{Tr}_\text{A}\Big([\hat{R}_{\text{A},i},\hat{R}_{\text{S},j}]\rho_\text{A}\Big) \ 
[\hat{Q}_{\text{S},i},\hat{Q}_{\text{S},j}]
\neq0
\ee
In order for \eqref{GenCond} to be non-zero, a Hamiltonian of the form \eqref{HsaDecomp} must have a pair of terms whose system parts do not commute and whose ancilla parts do not commute on average.

Thus rapid repeated interactions with ancillae under an arbitrary Hamiltonian \eqref{HsaDecomp} will in general be able to purify at leading order in decoherence effects. In section \ref{Examples}, we will show some simple non-product interactions that can purify at leading order. Conversely, we will also show some remarkable common types of non-product coupling that, nevertheless, cannot purify at leading order due to cancellations within \eqref{SumProduct}.

Note that while the above analysis only considers the specific form of the update map given by \eqref{TimelessPhi}, we can extend these results to a much wider class of update maps by making use of the results described at the end of Section \ref{PStrength}.

\subsection{Time-dependent interactions}
Additionally our analysis easily extends to include cases of ancillary bombardment where the Hamiltonian is explicitly time dependent. The dissipation effects in this scenario were analyzed in \cite{Grimmer2016a}. In particular they considered the Hamiltonian to be of the form
\bel{HamHam}
\hat{H}_{\delta t}(t)
=\hat{H}_\text{S}\otimes\boldsymbol{\hat{1}}
+\boldsymbol{\hat{1}}\otimes \hat{H}_\text{A}
+\hat{H}_\text{SA}(t/\delta t).
\ee
The update map for such an interaction is given by
\bel{PhiDef}
\phi(\delta t)[\rho_\text{S}]
=\text{Tr}_\text{A}\Big(U_{\delta t}(\delta t)(\rho_\text{S}\otimes \rho_{\text{A}}) U_{\delta t}(\delta t)^\dagger\Big)
\ee
where $U_{\delta t}(t)$ is the unitary transformation,
\be
U_{\delta t}(t)
=\mathcal{T}\exp\Big(\int_0^t \d\tau \, \hat{H}_{\delta t}(\tau)\Big)
\ee
and  $\mathcal{T}$ is the time-ordering operation. This unitary transformation is generated by a time-dependent Hamiltonian, $\hat{H}_{\delta t}(t)$. From \cite{Grimmer2016a}, we can compute the effect of $\mathcal{L}_1$ on the maximally mixed state, \eqref{L1oI}, as
\bel{Readoff}
\mathcal{L}_1[I]
=\frac{1}{2}
\Big(\frac{-\ii}{\hbar}\Big)^2
\text{Tr}_\text{A}\Big(
\big[G_0(\hat{H}_\text{SA}),[G_0(\hat{H}_\text{SA}),I\otimes \rho_\text{A}]\big]\Big)
\ee
where 
\bel{G0def} 
G_0(\hat{H}_\text{SA})\coloneqq
\int_0^1 \hat{H}_\text{SA}(\xi) \, d\xi 
\ee
is the unweighted time average of the interaction Hamiltonian. 

Thus we can see that the ability of an interaction Hamiltonian to purify at leading order in decoherence effects only depends on its time average. Thus a time dependent interaction can purify at leading order if and only if its time average can.

\section{Examples}\label{Examples}
In this section, we investigate several specific interaction Hamiltonians in light of the necessary and sufficient condition to purify at leading order which we described in the previous section. Namely, that when written as a sum of tensor products, \eqref{HsaDecomp}, it must satisfy \eqref{GenCond}. 

\subsection{Isotropic spin coupling (\texorpdfstring{$\bm{\hat{\sigma}}_\text{S}\cdot\bm{\hat{\sigma}}_\text{A}$}{text})}
As an example of an interaction capable of purifying at leading order, we consider the isotropic spin-spin interactions, 
\be
\hat{H}_\text{SA}
=\hbar \, J \, \bm{\hat{\sigma}}_\text{S}\cdot\bm{\hat{\sigma}}_\text{A}
=\hbar \, J \, \hat{\sigma}_\text{S}{}^j\otimes\hat{\sigma}_\text{A}{}_j
\ee
where we use Einstein's summation notation of implicitly summing over all repeated indices.

From \eqref{SumProduct} we can compute the effect of $\mathcal{L}_1$ on the maximally mixed state as,
\begin{align}
\mathcal{L}_1[I]
&\nonumber
=\frac{1}{2}\Big(\frac{-\ii}{\hbar}\Big)^2 (\hbar J)^2 \ 
\big\langle[\hat{\sigma}_\text{A}{}_i,\hat{\sigma}_\text{A}{}_j]\big\rangle \ [\hat{\sigma}_\text{S}{}^i,\hat{\sigma}_\text{S}{}^j]\\
&=4 \, J^2 \ 
\langle \bm{\hat{\sigma}}_\text{A}\rangle \,\cdot\, \bm{\hat{\sigma}}_\text{S}
\label{sig-sig}
\end{align}
where, for convenience, we have introduced the notation $\langle\,\cdot\,\rangle=\text{Tr}_\text{A}(\,\cdot\,\rho_\text{A})$.

In terms of Bloch vectors, eq. \eqref{sig-sig} expresses the intuitive result that the maximally mixed state, $\bm{a}_\text{S}=0$, is moved in the direction of the ancilla's Bloch vector, $\bm{a}_\text{A}=\langle \bm{\hat{\sigma}}_\text{A}\rangle$. Thus unless the ancillae are in the maximally mixed state there will be purification effects in the dynamics.

\subsection{Qubit-Harmonic Oscillator coupling}
We find another example of an interaction Hamiltonian that can purify by considering a qubit which repeatedly interacts with sequence of harmonic oscillators via the interaction Hamiltonian
\be
\hat{H}_\text{SA}=
\hbar\omega \, (\hat{x}\otimes\hat{\sigma}_x
+\hat{p}\otimes\hat{\sigma}_y)
\ee
where $\hat{x}=(\hat{a}+\hat{a}^\dagger)/2$ and $\hat{p}=\ii(\hat{a}-\hat{a}^\dagger)/2$ are quadrature operators for a harmonic oscillator. From \eqref{SumProduct} we can compute the effect of $\mathcal{L}_1$ on the maximally mixed state as,
\begin{align}
\mathcal{L}_1[I]
&\nonumber
=\Big(\frac{-\ii}{\hbar}\Big)^2 (\hbar \omega)^2 \ 
\big\langle[\hat{x},\hat{p}]\big\rangle \ [\hat{\sigma}_x,\hat{\sigma}_y]\\
&=2 \, \omega^2 \ \hat{\sigma}_z.
\end{align}
Thus the maximally mixed state is initially polarized in the $z$ direction under this interaction regardless of the state of the harmonic oscillator ancillae. 
  This type of interaction can, in principle, be implemented in superconducting circuits \cite{sufluxqubit}, achieving fast switching times in the ultra strong switchable coupling regime \cite{PeroPadre}.

\subsection{Vector-vector Couplings}\label{VectorVector}
As discussed in section \ref{TPI} rapidly interacting with an ancilla via a tensor product of two scalar observables, $\hat{H}_\text{SA}=\hat{Q}_\text{S}\otimes \hat{R}_\text{A}$, cannot purify at leading order. A natural generalization of this coupling is to instead couple two vector observables component-wise (through their dot product) as, 
\bel{VectorCoupling}
\hat{H}_\text{SA}
=\bm{\hat{V}}_\text{S}\cdot\bm{\hat{W}}_\text{A}
\coloneqq\hat{V}_\text{S}{}^j\otimes\hat{W}_\text{A}{}_j.
\ee

From \eqref{SumProduct}, the effect of $\mathcal{L}_1$ on the maximally mixed state is
\be
\mathcal{L}_1[I]
=\frac{1}{2}\Big(\frac{-\ii}{\hbar}\Big)^2 \ 
\big\langle[\hat{W}_\text{A}{}_i,\hat{W}_\text{A}{}_j]\big\rangle \ [\hat{V}_\text{S}{}^i,\hat{V}_\text{S}{}^j].
\ee
Thus, for repeated interactions under \eqref{VectorCoupling} to purify, the components of $\bm{\hat{V}}$ must not commute amongst themselves, and the components of $\bm{\hat{W}}$ must not either.

Many common vector observables such as $\bm{\hat{x}}$, $\bm{\hat{p}}$, $\bm{\hat{E}}(\bm{x})$, and $\bm{\hat{B}}(\bm{x})$, do not pass this test, while others such as $\bm{\hat{L}}$ and $\bm{\hat{\sigma}}$ do. Thus vector-vector couplings involving any of $\bm{\hat{x}}$, $\bm{\hat{p}}$, $\bm{\hat{E}}(\bm{x})$, or $\bm{\hat{B}}(\bm{x})$ can not purify whereas couplings involving $\bm{\hat{L}}$ or $\bm{\hat{\sigma}}$ potentially can depending on what they are coupled to. 

From this we can generalize further to the case of two vector fields coupled component-wise throughout all of space as,
\begin{align}\label{VectorFieldCoupling}
H_\text{SA}
=\int \d\bm{x} \ \bm{\hat{V}}_\text{S}(\bm{x})\cdot\bm{\hat{W}}_\text{A}(\bm{x})
=\int \d\bm{x} \ \hat{V}_\text{S}{}^j(\bm{x})\otimes\hat{W}_\text{A}{}_j(\bm{x}).
\end{align}
A necessary condition for repeated interaction this type of Hamiltonian to purify is that at least one of the following two conditions holds:
\begin{enumerate}
\item Neither $\bm{\hat{V}}(\bm{x})$ nor $\bm{\hat{W}}(\bm{x})$ is microcausal. (Recall that an observable $\hat{\bm{X}}$ is microcausal if \mbox{$[\hat{X}_i(\bm{x}),\hat{X}_j(\bm{x}')]$} only has support on $\bm{x}=\bm{x}'$).
\item Neither $\bm{\hat{V}}(\bm{x})$ nor $\bm{\hat{W}}(\bm{x})$ has its components commute locally amongst themselves, i.e. \mbox{$[\hat{X}_i(\bm{x}),\hat{X}_j(\bm{x})]\neq0$}.
\end{enumerate}
To see that this is the case, we compute the effect of $\mathcal{L}_1$ on the maximally mixed state from \eqref{SumProduct} as,
\begin{align}\label{{L1oIVW}}
&\mathcal{L}_1[I]\\
&\nonumber
=\Big(\frac{-\ii}{\hbar}\Big)^2\!\!\!\int\!\!\! \d\bm{x}\!\!\int\!\!\! d\bm{x'} \, 
\big\langle [\hat{W}_i(\bm{x}),\hat{W}_j(\bm{x'})]\big\rangle \, 
[\hat{V}^i(\bm{x}),\hat{V}^j(\bm{x'})]
\end{align}
If one of $\bm{V}(\bm{x})$ or $\bm{W}(\bm{x})$ is microcausal then the integral's domain can be reduced to the $\bm{x}=\bm{x}'$ region. From there, whichever of $\bm{V}(\bm{x})$ or $\bm{W}(\bm{x})$ has its components locally commuting causes the integrand to vanish. Thus such interactions cannot purify at leading order.

\subsection{Light-matter Interaction}
Let us now focus on a concrete relevant model used in quantum optics: we will analyze the ability of the light-matter interaction to purify in the context of rapid repeated interactions.

Let us consider an atom interacting with a second-quantized electromagnetic field. Let us take the atom as the target system, S, and the field as the ancilla, A, to which the system is repeatedly coupled. Physically, one can imagine atoms bombarded by pulses of light.

We begin by showing that any single multipolar coupling of the electric field to an atom cannot purify at leading order in rapid repeated interactions. 

First, we consider the electric dipole interaction given by 
\begin{align}
\hat{H}_\text{SA}
=q \, \hat{x}^j\hat{E}_j
&\nonumber
=\int \d\bm{x} \ q \, \hat{x}^j \, \ket{\bm{x}}\!\bra{\bm{x}}\otimes\hat{E}_j(\bm{x})\\
&=\int \d\bm{x} \ \hat{d}^j(\bm{x})\otimes\hat{E}_j(\bm{x})
\end{align}
where $\hat{d}^j(\bm{x})=q \, x^j \, \ket{\bm{x}}\!\bra{\bm{x}}$ is the dipole moment operator at a position $\bm{x}$ \cite{Scully:1997,Pozas2016a}. 

In this form, the interaction is written as the coupling of two vector fields throughout all of space. This is the scenario that was analyzed at the end of the previous section. It is enough to note that the electric field is microcausal and the components of $\bm{\hat{d}}(\bm{x})$ commute amongst themselves locally (in fact, both observables have both properties) to conclude that the electric dipole interaction cannot purify at leading order on its own.

Similarly, if we consider the electric quadrupole coupling given by
\be
H_\text{SA}
=q \, \hat{x}^i\hat{x}^j\nabla_i\hat{E}_j
=\!\!\int\!\! \d\bm{x} \ \hat{Q}^i{}^j(\bm{x})\otimes\bm{\nabla}_i\hat{E}_j(\bm{x})
\ee
where $\hat{Q}^i{}^j(\bm{x})=q \, x^i x^j \ket{\bm{x}}\!\bra{\bm{x}}$ is the quadrupole moment operator at a position $\bm{x}$, we find that it cannot purify at leading order. This is again because $\bm{\nabla}_i\hat{E}_j(\bm{x})$ is microcausal and the components of $\hat{Q}^i{}^j(\bm{x})$ commute amongst themselves locally.

Similarly, every higher multipolar electric coupling cannot purify on its own at leading order since higher derivatives of the electric field remain microcausal and the components of the higher moment operators always commute amongst themselves locally.

A similar analysis can be carried out with the magnetic dipole interaction given by
\bel{MD1}
\hat{H}_\text{SA}=\frac{q}{2m}\{\hat{L}^k,\hat{B}_k\}
=\int d\bm{x} \ \hat{\mu}^k(\bm{x})\otimes\hat{B}_k(\bm{x})
\ee
where $\hat{\mu}^k(\bm{x})=(q/2m)(\hat{L}^k\ket{\bm{x}}\!\bra{\bm{x}}+\ket{\bm{x}}\!\bra{\bm{x}}\hat{L}^k)$ is the magnetic dipole operator at a position $\bm{x}$. The Hamitonian \eqref{MD1} is again the coupling of two vector fields throughout all of space. We conclude as before that the magnetic dipole interaction cannot purify at leading order since the magnetic field is both microcausal and has its components commute amongst themselves locally.

Furthermore, linear combinations of different electric multipole couplings cannot purify at leading order either. For example consider the combination of electric dipole and electric quadrupole interactions
\begin{align}
H_\text{SA}
&=q \, \hat{x}^k\hat{E}_k
+q \, \hat{x}^i\hat{x}^j\nabla_i\hat{E}_j\\
&\nonumber
=\!\!\int\!\! \d\bm{x} \ 
\Big(\hat{d}^k(\bm{x})\otimes\hat{E}_k(\bm{x})
+\hat{Q}^i{}^j(\bm{x})\otimes\bm{\nabla}_i\hat{E}_j(\bm{x})\Big).
\end{align}
In computing the effect of $\mathcal{L}_1$ on the maximally mixed state, \eqref{SumProduct}, the cross terms within the dipole coupling will vanish, as will the cross terms within the quadrupole coupling. Only the cross terms between the two couplings remain, yielding,
\begin{align}
&\mathcal{L}_1[I]\\
&\nonumber
=\Big(\frac{-\ii}{\hbar}\Big)^2\!\!\!\int\!\!\! \d\bm{x}\!\!\int\!\!\! \d\bm{x'} \, 
\big\langle [\hat{E}_k(\bm{x}),\nabla_i\hat{E}_j(\bm{x'})]\big\rangle \, 
[\hat{d}^k(\bm{x}),\hat{Q}^i{}^j(\bm{x'})].
\end{align}
However, this too vanishes since,
\be
[\hat{d}^k(\bm{x}),\hat{Q}^i{}^j(\bm{x'})]=0
\ee
for all $\bm{x}$ and $\bm{x}'$. This can be easily seen by noting that both $\hat{d}^k(\bm{x})$ and $\hat{Q}^i{}^j(\bm{x})$ are diagonal in the position basis. 

In the same fashion, any combination of any electric multipolar couplings will not be able to purify at leading order. Thus rapid repeated light-matter interactions where matter couples only to the electric field are unable to purify at leading order.

Thus if we have any hope of purifying at leading order we must involve the magnetic field. A first very simple combination of electric and magnetic couplings that we can consider is the combination of the electric dipole and magnetic dipole couplings:
\begin{align}
H_\text{SA}
&=q \, \hat{x}^j\hat{E}_j
+\frac{q}{2m}\{\hat{L}^k,\hat{B}_k\}\\
&\nonumber
=\int \d\bm{x} \ \Big( \hat{d}^j(\bm{x})\otimes\hat{E}_j(\bm{x})
+\hat{\mu}^k(\bm{x})\otimes\hat{B}_k(\bm{x})\Big).
\end{align}
This interaction Hamiltonian satisfies the necessary condition for purification to appear at leading order, as discussed in previous sections. Namely, $[\hat{d}^j(\bm{x}),\hat{\mu}^k(\bm{x}')]\neq0$ and $[\hat{E}_j(\bm{x}),\hat{B}_k(\bm{x}')]\neq0$.

As above the cross terms within each of the electric and magnetic couplings will vanish and only the commutators mixing the electric and magnetic field will survive. Computing the effect of $\mathcal{L}_1$ on the maximally mixed states, \eqref{SumProduct}, yields,
\be
\mathcal{L}_1[I]
\nonumber
=\Big(\frac{-\ii}{\hbar}\Big)^2\!\!\!\int\!\!\! \d\bm{x}\!\!\int\!\!\! d\bm{x'} \, 
\big\langle [\hat{E}_j(\bm{x}),\hat{B}_k(\bm{x'})]\big\rangle \, 
[\hat{d}^j(\bm{x}),\hat{\mu}^k(\bm{x'})]
\ee
This integrand is non-zero but, remarkably, the integral over $\bm{x}$ and $\bm{x}'$ vanishes. The mechanism for this cancellation is particularly interesting, and is discussed in detail in Appendix \ref{EDiMDi}. Therefore, despite the fact that the coupling satisfies the necessary condition for purification discussed in section \ref{NPIH}, rapid repeated interactions which involve both the electric and magnetic dipole couplings cannot purify at leading order.

This result is easily extended to more general light-matter couplings. For example, in the case of the more physically relevant combination of both the electric quadrupole and magnetic couplings
\begin{align}
H_\text{SA}
&=q \, \hat{x}^i\hat{x}^j\nabla_i\hat{E}_j
+\frac{q}{2m}\{\hat{L}^k,\hat{B}_k\}\\
&\nonumber
=\int \d\bm{x} \ \Big( \hat{Q}^i{}^j(\bm{x})\otimes\nabla_i\hat{E}_j(\bm{x})
+\hat{\mu}^k(\bm{x})\otimes\hat{B}_k(\bm{x})\Big).
\end{align}
we find a similar cancellation that yields no purification at leading order. Any higher order electric couplings will exhibit the same cancellation as will the combination of several electric multipolar moments along with the magnetic dipole moment. 

Summarizing, we have proven that the most common models of light-matter interactions employed in quantum optics \cite{Scully:1997}---i.e., those involving any combination of electric multipolar couplings and the magnetic dipole coupling---cannot purify at leading order under rapid repeated interactions. 

\section{Conclusion}
We analyzed the ability of rapid repeated interactions to purify a quantum system. In particular, we considered the formalism developed in \cite{Layden:2015b,Grimmer2016a}, where a quantum system evolves (in discrete time steps of duration $\delta t$) under the repeated application of a quantum channel. 

We have studied and characterized the strength of purification effects of these rapid repeated interactions, namely, at what order in $\delta t$ the dynamics can lead to purification. We have shown that, perhaps contrary to intuition, the purification strength cannot be increased by combining different rapid repeated interaction dynamics by composition or convex combination.

After this general study, we have investigated in-depth the purifying power of a particularly relevant scenario that we called \textit{ancillary bombardment}. In this scenario a quantum system is bombarded by a sequence of ancillae, undergoing a brief unitary interaction with each of them. For instance, one can think of an atom interacting with its environment, under the assumption that it repeatedly interacts unitarily with its individual constituents. Another example of such a scenario would be a laboratory system that is repeatedly measured by probes.

We have shown that simple interaction Hamiltonians (including some considered in previous literature on rapid repeated interactions \cite{CM}) cannot purify at leading order if their interaction strength remains finite. Furthermore, we have shown that for an ancillary bombardment to purify at leading order it must be mediated by a sufficiently complicated Hamiltonian. Specifically, an interaction consisting of the tensor products of two scalar observables will not purify at leading order.

We have found necessary and sufficient conditions for a ancillary bombardment to purify a quantum system. We studied what kinds of couplings satisfy them and what kind of couplings do not. For illustration, we have shown how an isotropic spin-spin coupling, as well as a specific experimentally feasible interaction of a qubit with a harmonic oscillator can purify at leading order under ancillary bombardment.

Furthermore, we have paid special attention to the case of couplings of system observables to vector fields, and in particular the case of the multipole moments of an atom coupled to the fully quantized electromagnetic field (EM).

For the case of interaction with relativistic quantum fields (such as the EM field) we have found necessary conditions for purification involving the microcausality of the theory.

Remarkably, we have shown that any combinations of electric multipole couplings and the magnetic dipole coupling cannot purify at leading order under repeated interaction. This casts fundamental doubt on the ability of simple quantum optical setups to increase the purity of atomic qubits under fast interaction.

These results may perhaps be relevant to the field of algorithmic cooling  and can be used to design setups to prolong the life of quantum coherence through a controlled exposure to an environment. The particular implications of these results in quantum thermodynamics are intriguing and will be analyzed elsewhere.

\section*{Acknowledgements}

This work was supported in part by the Natural Sciences and Engineering Research Council of Canada through the NSERC Discovery programme.

\appendix
\section{Necessity of Non-unitality for Purification}\label{Necessity}
In this appendix we reproduce (in our notation) a proof given in \cite{Lidar2006}. Specifically we prove that, for a finite dimensional systems, in order for the dynamics generated by a Liouvillian, $\mathcal{L}$, to cause purification it is necessary that the dynamics are not unital, that is, $\mathcal{L}[I]\neq0$, where and $I$ is the \mbox{$d$-dimensional} identity matrix. In order to show this we derive the inequality,
\be
\frac{\d}{\d t}\mathcal{P}(\rho)
=\frac{\d}{\d t}\text{Tr} \, (\rho{}^2)
\leq\text{Tr} \, \big(\mathcal{L}[I] \, \rho^2\big).
\ee
from which our claim follows directly.

We first write the Liouvillian in a standard form called the Lindblad form, that is,
\bel{LindbladForm}
\mathcal{L}[\rho] 
=\frac{-\ii}{\hbar}\big[\hat{H},\rho\big]
+\sum_n \Gamma_n \, \Big(\hat{F}_n\rho \hat{F}_n^\dagger-\frac{1}{2}\big\{\hat{F}_n^\dagger \hat{F}_n,\rho\big\}\Big),
\ee
where $\hat{H}$ is a Hermitian operator, $\hat{F}_n$ are operators, and $\Gamma_n$ are non-negative numbers. The operator $\hat{H}$ is the effective Hamiltonian of the dyanamics and is said to generate the unitary part of the dynamics. The operators $\hat{F}_n$ are the dynamics decoherence modes and $\Gamma_n$ are their respective decoherence rates.  Any Liouvillian can be written in this form \cite{Lindblad}. Note that the effect of the dynamics on the maximally mixed state is, 
\bel{LoI}
\mathcal{L}[I] 
=\sum_n \Gamma_n \,  [F_n,F_n^\dagger].
\ee

Using the cyclic property of trace we find the rate of change of systems purity is,
\bel{PurityRate}
\frac{\d}{\d t}\mathcal{P}(\rho)
=\frac{\d}{\d t}\text{Tr}(\rho^2)
=2 \, \text{Tr} \, \big(\mathcal{L}[\rho] \, \rho\big).
\ee
The unitary part of the dynamics does not change the purity, as expected, since 
\bel{UnitaryCancels}
\text{Tr}\big([H,\rho] \, \rho\big)
=\text{Tr}\big(H \, [\rho,\rho]\big)
=0.
\ee
Thus we can focus our attention on the decoherence modes. Using \eqref{PurityRate}, \eqref{UnitaryCancels}, and the cyclic property of trace we have
\begin{align}
\frac{\d}{\d t}\mathcal{P}(\rho)
&=2 \, \text{Tr} \, \big(\mathcal{L}[\rho] \, \rho\big)\\
&\nonumber
=2 \, \text{Tr} \, \Big(\sum_n \Gamma_n \, \big(F_n\rho F_n^\dagger-\{F_n^\dagger F_n,\rho\}/2\big)\rho\Big)\\
&\nonumber
=\sum_n \Gamma_n \, 2 \, \text{Tr} \, \big(
F_n\rho F_n^\dagger\rho
-F_n^\dagger F_n\rho^2\big).
\end{align}
For Hermitian $\rho$, we have the identity,
\be
2 \, \text{Tr}\big(A\rho A^\dagger\!\rho
-\! A^\dagger\!A\rho^2\big)
\!=\!\text{Tr}\big([A,A^\dagger]\rho^2\!
-\![A,\rho]^\dagger [A,\rho]\big),
\ee
which yields,
\begin{align}
\frac{\d}{\d t}\mathcal{P}(\rho)
&=\sum_n \Gamma_n \, \text{Tr} \, \big(
[F_n,F_n^\dagger]\rho^2
-[F_n,\rho]^\dagger [F_n,\rho]\big)\\
&\nonumber
=\text{Tr} \, \big(\mathcal{L}[I] \, \rho^2\big)
-\sum_n \Gamma_n \, \text{Tr} \, \big([F_n,\rho]^\dagger [F_n,\rho]\big)
\end{align}
where we have made use of \eqref{LoI} to identify $\mathcal{L}[I]$ in the first term. Since the second term is manifestly negative we have the inequality,
\begin{align}
\frac{\d}{\d t}\mathcal{P}(\rho)
\leq\text{Tr} \, \big(\mathcal{L}[I] \, \rho^2\big)
\end{align}
claimed in \eqref{PurityRateBound}. If $\mathcal{L}[I]=0$ then the dynamics will either maintain or decrease the purity of any state.

In \cite{Lidar2006}, this proof is shown to hold as well for infinite dimensional systems following some assumptions on the decoherence modes. In particular, it holds if all $\hat{F}_n$ are bounded.

\section{Concatenation And Convex Combinations}\label{CaCC}
In this appendix, we prove that constructing a new map by taking either concatenations or convex combinations of different maps cannot lower the resultant purification order, defined in \eqref{PurificationOrder}, below those of the original maps.

\subsection{Concatenation}
Suppose we have an update map $\phi(\delta t)$ that is the concatenation of two maps $\chi^{(1)}(\delta t)$ and $\chi^{(2)}(\delta t)$:
\be
\phi(\delta t)
=\chi^{(1)}(\delta t)\chi^{(2)}(\delta t).
\ee
We can take these two maps to have series expansions about $\delta t=0$  
\begin{align}
\label{Chi1Series}
\chi^{(1)}(\delta t)
&=\openone
+\delta t \, \chi^{(1)}_1
+\delta t^2 \, \chi^{(1)}_2
+\delta t^3 \, \chi^{(1)}_3
+\dots\\
\label{Chi2Series}
\chi^{(2)}(\delta t)
&=\openone
+\delta t \, \chi^{(2)}_1
+\delta t^2 \, \chi^{(2)}_2
+\delta t^3 \, \chi^{(2)}_3
+\dots \, 
\end{align}
and define
$m_1=\text{ord}(\chi^{(1)})$ and $m_2=\text{ord}(\chi^{(2)})$ to be the purification orders of $\chi^{(1)}(\delta t)$ and $\chi^{(2)}(\delta t)$ respectively, with  $m=\text{min}(m_1,m_2)$. Recall that a map's purification order is defined in terms of its interpolation scheme. Converting this into terms of the update maps we have \eqref{Offset} such that,
\begin{align}
&\chi^{(1)}[I]=I+\mathcal{O}(\delta t^{m_1+1})\\
&\nonumber
\chi^{(2)}[I]=I+\mathcal{O}(\delta t^{m_2+1}).
\end{align}
Thus we have
\begin{align}\label{LesserOrder1}
&\chi^{(1)}_1[I]=\dots=\chi^{(1)}_{m}[I]=0\\
&\nonumber
\chi^{(2)}_1[I]=\dots=\chi^{(2)}_{m}[I]=0.
\end{align}
Evaluating $\phi(\delta t)$ on the maximally mixed state yields
\begin{align}
&\phi(\delta t)
=\chi^{(1)}(\delta t)\chi^{(2)}(\delta t)\\
&\nonumber
=\big(\openone
+\sum_{k=1}^\infty
\delta t^k \, \chi^{(1)}_k\big)\big(\openone
+\sum_{n=1}^\infty
\delta t^n \, \chi^{(2)}_n\big)[I]\\
&\nonumber
=\big(\openone
+\sum_{k=1}^\infty
\delta t^k \, \chi^{(1)}_k\big)\big(I
+\sum_{n=1}^\infty
\delta t^n \, \chi^{(2)}_n[I]\big)\\
&\nonumber
=I
+\sum_{k=1}^\infty
\delta t^k \, \chi^{(1)}_k[I]
+\sum_{n=1}^\infty
\delta t^n \, \chi^{(2)}_n[I]
+\sum_{k=1}^\infty\sum_{n=1}^\infty
\delta t^{k+n} \, \chi^{(1)}_k[\chi^{(2)}_n[I]]\\
&\nonumber
=I
+\sum_{k=m+1}^\infty
\delta t^k \, \chi^{(1)}_k[I]
+\sum_{n=m+1}^\infty
\delta t^n \, \chi^{(2)}_n[I]\\
&+\sum_{k=1}^\infty\sum_{n=m+1}^\infty
\delta t^{k+n} \, \chi^{(1)}_k[\chi^{(2)}_n[I]]
\end{align}
where we have used \eqref{LesserOrder1} to drop terms from the sums. From this we can see that any non-unital effects in $\phi(\delta t)$ appear at at least order $m+1$ and thus $\text{ord}(\phi)\geq m$.

By applying this proof repeatedly one can conclude that if $\phi(\delta t)$ is a concatenation of a finite number of maps as,
\bel{Concatenation}
\phi(\delta t)=
\chi^{(1)}(\delta t) \, \chi^{(2)}(\delta t) \, \dots \, \chi^{(N)}(\delta t),
\ee
then 
\be
\text{ord}(\phi)\geq\text{min}\{\text{ord}(\chi^{(n)})\}
\ee
as claimed.

\subsection{Convex Combinations}
Suppose we have an update map $\phi(\delta t)$ which is a convex combination of maps as,
\bel{ConvexCombo}
\phi(\delta t)=
\sum_k p_k \, \psi^{(k)}(\delta t),
\ee
with $\sum_k p_k =1$. We can take these maps to have series expansions about $\delta t=0$ as,
\bel{PsikSeries}
\psi^{(k)}(\delta t)
=\openone
+\delta t \, \psi^{(k)}_1
+\delta t^2 \, \psi^{(k)}_2
+\delta t^3 \, \psi^{(k)}_3
+\dots \, .
\ee
Let $m_k=\text{ord}(\psi^{(k)})$ be the purification orders of $\psi^{(k)}(\delta t)$ and $m=\text{min}\{m_k\}$. Recall that a map's purification order is defined in terms of its interpolation scheme. Converting this into terms of the update maps we have \eqref{Offset} such that,
\be
\psi^{(k)}[I]=I+\mathcal{O}(\delta t^{m_k+1})
\ee
and thus for every $k$,
\bel{LesserOrder2}
\psi^{(k)}_1[I]=\dots=\psi^{(k)}_m[I]=0.
\ee
Evaluating $\phi(\delta t)$ on the maximally mixed state yields
\begin{align}
\phi(\delta t)[I]
&=\sum_k p_k \, \psi^{(k)}(\delta t)[I]\\
&\nonumber
=\sum_k p_k \, \big(\openone
+\sum_{n=1}^\infty
\delta t^n \, \psi^{(k)}_n\big)[I]\\
&\nonumber
=\big(\openone
+\sum_{n=1}^\infty
\delta t^n \, \sum_k p_k \, \psi^{(k)}_n\big)[I]\\
&\nonumber
=I+\sum_{n=1}^\infty \delta t^n \,
\sum_k p_k \, \psi^{(k)}_n[I]\\
&\nonumber
=I+\sum_{n=m+1}^\infty \delta t^n \,
\sum_k p_k \, \psi^{(k)}_n[I]
\end{align}
where in the last step we have used \eqref{LesserOrder2} to drop terms from the sums. From this we can see that any non-unital effects in $\phi(\delta t)$ appear at at least order $m+1$ and thus $\text{ord}(\phi)\geq m$ as claimed.

\section{Calculation of Purification Orders}\label{TPCalc}
In this appendix we find the leading order purification effects in the ancillary bombardment scenario discussed in section \ref{AB}, for several different interaction Hamiltonians.

\subsection{History Reduction}
In the general case, the system and ancilla interact via the joint Hamiltonian,
\bel{HSATP}
\hat{H}
=\hat{H}_\text{S}\otimes\openone
+\openone\otimes \hat{H}_\text{A}
+\hat{H}_\text{SA}.
\ee

To find the leading order purification effects we compute $\mathcal{L}_{\delta t}[I]$ order by order until we find the first non-zero contribution. However due to the recursive structure of the coefficient maps in \eqref{LdtSeries} we can simply look for the smallest $m$ such that $\phi_m$ that moves the identity, $\phi_m[I]\neq I$. These maps are given by the partial trace of $m$ nested commutations with $\hat{H}$ applied to $I\otimes \rho_\text{A}$. For instance,
\be
\phi_4[I]
=\frac{1}{4!}\Big(\frac{-\ii}{\hbar}\Big)^4
\text{Tr}_\text{A}\Big(
[\hat{H},[\hat{H},[\hat{H},[\hat{H},I\otimes \rho_\text{A}]]]]\Big).
\ee
By the linearity of the commutator, these computations involve all possible ways of picking one of the three terms from \eqref{HSATP} for each of the $m$ commutators. In a sum over histories sense, $\phi_m$ involves all possible ways of the system and ancilla meeting $m$ times, each time selecting one of $\hat{H}_\text{SA}$, $\hat{H}_\text{S}\otimes\openone$, or
$\openone\otimes \hat{H}_\text{A}$ to evolve under. In human terms, each day they may either interact with the wider world or stay home and reflect on their lives.

In order to simplify the following computations we first work out some immediate reductions that happen when choosing either of the free Hamiltonians for either the inner most or outer most commutator.

First we see that picking the ancilla's free Hamiltonian for the outermost commutator causes a history's contribution to vanish. This follows directly from the cyclic property of partial trace, namely
\be
\text{Tr}_\text{A}\big(
(\openone\otimes\hat{H}_\text{A})\hat{Z}_\text{SA}\big)
=\text{Tr}_\text{A}\big(
\hat{Z}_\text{SA}(\openone\otimes\hat{H}_\text{A})\big).
\ee
for any $\hat{Z}_\text{SA}$ such that
\bel{CyclicPropertyOfPartialTrace}
\text{Tr}_\text{A}\big(
[\openone\otimes\hat{H}_\text{A},\hat{Z}_\text{SA}]\big)=0.
\ee
Thus choosing $\hat{H}_\text{A}$ for the outer most commutator yields
\begin{align}
\text{Tr}_\text{A}\Big(
[\openone\otimes\hat{H}_\text{A},[\hat{H},[\dots,[\hat{H},I\otimes \rho_\text{A}]]]]\Big)=0.
\end{align}
Note, if the ancillae are infinite dimensional then the above calculation requires that all relevant ancilla observables are trace class.

Additionally, if one selects the system free Hamiltonian for the inner most commutator one finds,
\begin{align}
\text{Tr}_\text{A}\Big(
[\hat{H},[\dots,[\hat{H},[\hat{H}_\text{S}\otimes\openone,I\otimes \rho_\text{A}]]]]\Big)=0
\end{align}
since $\hat{H}_\text{S}\otimes\openone$ and $I\otimes \rho_\text{A}$ act on disjoint sectors of the Hilbert space.

On the other hand, if one selects the ancilla free Hamiltonian for the innermost commutator the result is expressible in terms of $\phi_{m-1}[I]$. Specifically one finds,
\begin{align}
&\text{Tr}_\text{A}\Big(
[\hat{H},[\dots,[\hat{H},[\openone\otimes\hat{H}_\text{A},I\otimes \rho_\text{A}]]]]\Big)\\
&\nonumber
=\text{Tr}_\text{A}\Big(
[\hat{H},[\dots,[\hat{H},I\otimes[\hat{H}_\text{A},\rho_\text{A}]]]]\Big)\\
&\nonumber
\sim\phi_{m-1}[I] \ \ \text{but with }\rho_\text{A}\to[\hat{H}_\text{A},\rho_\text{A}].
\end{align}
In particular, if $\phi_{m-1}[I]=0$ for any initial ancilla state, then picking $H_\text{A}$ for the inner most commutator does not add anything to the final result.

Finally, if one chooses $\hat{H}_\text{S}$ for the outer most commutator we also find that the result is expressible in terms of $\phi_{m-1}[I]$. To show this we first realize that, when acting on a tensor product, the actions `to commute with $\hat{H}_\text{S}(\otimes\openone)$' and  `to take the partial trace over A' commute. Concretely,
\be
\text{Tr}_\text{A}\big(
[\hat{H}_\text{S}\otimes\openone,\hat{Z}_\text{SA}]\big)
=\Big[\hat{H}_\text{S},\text{Tr}_\text{A}\big(\hat{Z}_\text{SA}\big)\Big].
\ee
for any $\hat{Z}_\text{SA}$. By linearity of the commutator and of partial trace we need only consider the case when $\hat{Z}_\text{SA}$ is a tensor product. In this case we find 
\begin{align}
\text{Tr}_\text{A}\big(
[\hat{H}_\text{S}\otimes\openone,\hat{X}_\text{S}\otimes\hat{Y}_\text{A}]\big)
&=\text{Tr}_\text{A}\Big(\big[\hat{H}_\text{S},\hat{X}_\text{S}\big]\otimes\hat{Y}_\text{A}\Big)\\
&\nonumber
=\big[\hat{H}_\text{S},\hat{X}_\text{S}\big] \, 
\text{Tr}_\text{A}\big(\hat{Y}_\text{A}\big)\\
&\nonumber
=\Big[\hat{H}_\text{S},\hat{X}_\text{S} \, \text{Tr}_\text{A}\big(\hat{Y}_\text{A}\big)\Big]\\
&\nonumber
=\Big[\hat{H}_\text{S},\text{Tr}_\text{A}\big(\hat{X}_\text{S}\otimes\hat{Y}_\text{A}\big)\Big].
\end{align}
Thus choosing $\hat{H}_\text{S}$ for the outermost commutator results in an expression of the form
\begin{align}
&\text{Tr}_\text{A}\Big(
[\hat{H}_\text{S}\otimes\openone,[\hat{H},[\dots,[\hat{H},I\otimes \rho_\text{A}]]]]\Big)\\
&\nonumber
=\Big[\hat{H}_\text{S},
\text{Tr}_\text{A}\big([\hat{H},[\dots,[\hat{H},I\otimes \rho_\text{A}]]]\big)\Big]\\
&\nonumber
\sim\big[\hat{H}_\text{S},
\phi_{m-1}[I]\big].
\end{align}
In particular, if $\phi_{m-1}[I]=0$, then picking $H_\text{S}$ for the outermost commutator does not add anything to the final result.

Taking these four cases into account we have the result that if $\phi_{m-1}[I]=0$ for every $\rho_\text{A}$ then the innermost and outermost commutators are forced to be $\hat{H}_\text{SA}$.

\subsection{Tensor Product Interaction}
In this subsection we show that in the ancillary bombardment scenario discussed in section \ref{AB}, if the system and ancilla interact via a tensor product of scalar observables as,
\bel{HSATP}
\hat{H}
=\hat{H}_\text{S}\otimes\openone
+\openone\otimes \hat{H}_\text{A}
+\hat{Q}_\text{S}\otimes \hat{R}_\text{A},
\ee
then the leading order purification effects are given by
\be
\mathcal{L}_3[I]
=\frac{1}{12\hbar^4}
[\hat{Q}_\text{S},[\hat{H}_\text{S},\hat{Q}_\text{S}]] \, 
\text{Tr}_\text{A}\Big([\hat{R}_\text{A},[\hat{H}_\text{A},\hat{R}_\text{A}]]\rho_\text{A}\Big).    
\ee

Proceeding order by order we first use \eqref{TimelessPhi1}, and \eqref{CyclicPropertyOfPartialTrace} to compute,
\begin{align}
\phi_1[I]
&=\frac{-\ii}{\hbar} \, 
\text{Tr}_\text{A}\Big(
[\hat{H},I\otimes \rho_\text{A}]\Big)\\
&\nonumber
=0
\end{align}

Next using \eqref{TimelessPhi2} we have
\be
\phi_2[I]=\frac{1}{2!}\Big(\frac{-\ii}{\hbar}\Big)^2
\text{Tr}_\text{A}\Big(
[\hat{H},[\hat{H},I\otimes \rho_\text{A}]]\Big).
\ee
Recalling the result derived earlier in this appendix, we know that, since $\phi_1[I]=0$ for every $\rho_\text{A}$, we must select the interaction Hamiltonian in both the inner most and outer most commutators. Thus,
\be
\phi_2[I]
=\frac{1}{2!}\Big(\frac{-\ii}{\hbar}\Big)^2
\text{Tr}_\text{A}\Big(
[\hat{Q}_\text{S}\otimes\hat{R}_\text{A},[\hat{Q}_\text{S}\otimes\hat{R}_\text{A},I\otimes \rho_\text{A}]]\Big).
\ee
Computing this yields zero.

Pressing on, from \eqref{TimelessPhi3} we have
\be
\phi_3[I]
=\frac{1}{3!}\Big(\frac{-\ii}{\hbar}\Big)^3
\text{Tr}_\text{A}\Big(
[\hat{H},[\hat{H},[\hat{H},I\otimes \rho_\text{A}]]]\Big).
\ee
Again, since $\phi_2[I]=0$ for every $\rho_\text{A}$, all histories without an interaction at the start and end vanish. Thus
\be
\phi_3[I]
=\frac{1}{3!}\Big(\frac{-\ii}{\hbar}\Big)^3
\text{Tr}_\text{A}\Big(
[\hat{Q}_\text{S}\otimes\hat{R}_\text{A},[\hat{H},[\hat{Q}_\text{S}\otimes\hat{R}_\text{A},I\otimes \rho_\text{A}]]]\Big).
\ee
The $\hat{H}$ in this expression yields three terms, all of which vanish.

Finally, from \eqref{TimelessPhi4} we have
\be
\phi_4[I]
=\frac{1}{4!}\Big(\frac{-\ii}{\hbar}\Big)^4
\text{Tr}_\text{A}\Big(
[\hat{H},[\hat{H},[\hat{H},[\hat{H},I\otimes \rho_\text{A}]]]]\Big).
\ee
Once again, since $\phi_3[I]=0$ for every $\rho_\text{A}$ we have,
\be
\phi_4[I]
=\frac{1}{4!}\Big(\frac{-\ii}{\hbar}\Big)^4
\text{Tr}_\text{A}\Big(
[\hat{Q}_\text{S}\otimes\hat{R}_\text{A},[\hat{H},[\hat{H},[\hat{Q}_\text{S}\otimes\hat{R}_\text{A},I\otimes \rho_\text{A}]]]]\Big).
\ee
The two $\hat{H}$ in this expression yield nine terms to check. All of them vanish except for the two terms with the free Hamiltonians in the middle. Thus,
\begin{align}
&\phi_4[I]\\
&\nonumber
=\frac{1}{4! \, \hbar^4} \text{Tr}_\text{A}\Big(
[\hat{Q}_\text{S}\otimes\hat{R}_\text{A},
[\hat{H}_\text{S}\otimes\openone,
[\openone\otimes\hat{H}_\text{A},
[\hat{Q}_\text{S}\otimes\hat{R}_\text{A},
I\otimes\rho_\text{A}]]]] \Big)\\
&\nonumber
+\frac{1}{4!\, \hbar^4}\text{Tr}_\text{A}\Big(
[\hat{Q}_\text{S}\otimes\hat{R}_\text{A},
[\openone\otimes\hat{H}_\text{A},
[\hat{H}_\text{S}\otimes\openone,
[\hat{Q}_\text{S}\otimes\hat{R}_\text{A},
I\otimes\rho_\text{A}]]]] \Big)\\
&\nonumber
=\frac{1}{12\hbar^4}
[\hat{Q}_\text{S},[\hat{H}_\text{S},\hat{Q}_\text{S}]] \, 
\text{Tr}_\text{A}\Big([\hat{R}_\text{A},[\hat{H}_\text{A},\hat{R}_\text{A}]]\rho_\text{A}\Big).
\end{align}
Heuristically, in a sum over histories sense, this process involves the system and ancillae interacting with each other, then each evolving freely, and finally interacting again. Using \eqref{L3def} we find
\begin{align}
\mathcal{L}_3[I]
&=\frac{1}{12\hbar^4}
[\hat{Q}_\text{S},[\hat{H}_\text{S},\hat{Q}_\text{S}]] \, 
\text{Tr}_\text{A}\Big([\hat{R}_\text{A},[\hat{H}_\text{A},\hat{R}_\text{A}]]\rho_\text{A}\Big)    
\end{align}
as claimed.

\subsection{Non-tensor product interactions}
In this subsection we show that in the ancillary bombardment scenario discussed in section \ref{AB}, if the system and ancilla interact via a sum of tensor products as,
\bel{HSATP}
\hat{H}
=\hat{H}_\text{S}\otimes\openone
+\openone\otimes \hat{H}_\text{A}
+\hat{Q}_\text{S}\otimes \hat{R}_\text{A}
+\hat{S}_\text{S}\otimes \hat{T}_\text{A},
\ee
then the leading order purification effects are given by
\be
\mathcal{L}_1[I]
=\frac{1}{(\ii\hbar)^2}
[\hat{Q}_\text{S},\hat{S}_\text{S}] \ 
\text{Tr}_\text{A}\Big([\hat{R}_\text{A},\hat{T}_\text{A}]\rho_\text{A}\Big)..    
\ee

Proceeding order by order we first use \eqref{TimelessPhi1}, and \eqref{CyclicPropertyOfPartialTrace} to compute,
\begin{align}
&\phi_1[I]
=\frac{-\ii}{\hbar} \, 
\text{Tr}_\text{A}\Big(
[\hat{H},I\otimes \rho_\text{A}]\Big)\\
&\nonumber
=0
\end{align}
Next, from \eqref{TimelessPhi2} we have
\be
\phi_2[I]=\frac{1}{2!}\Big(\frac{-\ii}{\hbar}\Big)^2
\text{Tr}_\text{A}\Big(
[\hat{H},[\hat{H},I\otimes \rho_\text{A}]]\Big).
\ee
Recalling the result derived earlier in this appendix, we know that since $\phi_1[I]=0$ for every $\rho_\text{A}$, we must select the interaction Hamiltonian in both the inner most and outer most commutators. Thus,
\begin{align}
&\phi_2[I]
=\frac{1}{2!}\Big(\frac{-\ii}{\hbar}\Big)^2\\
\nonumber
&\text{Tr}_\text{A}\Big(
[\hat{Q}_\text{S}\otimes\hat{R}_\text{A}
+\hat{S}_\text{S}\otimes \hat{T}_\text{A}
,[\hat{Q}_\text{S}\otimes\hat{R}_\text{A}
+\hat{S}_\text{S}\otimes \hat{T}_\text{A},I\otimes \rho_\text{A}]]\Big).    
\end{align}
Computing this yields
\begin{align}
\mathcal{L}_1[I]
\nonumber
&=\frac{1}{2(\ii\hbar)^2}\text{Tr}_\text{A}\Big(
[\hat{Q}_\text{S}\otimes\hat{R}_\text{A},
[\hat{S}_\text{S}\otimes \hat{T}_\text{A},
I\otimes\rho_\text{A}]] \Big)\\
&\nonumber
+\frac{1}{2(\ii\hbar)^2}\text{Tr}_\text{A}\Big(
[\hat{S}_\text{S}\otimes \hat{T}_\text{A},
[\hat{Q}_\text{S}\otimes\hat{R}_\text{A},
I\otimes\rho_\text{A}]] \Big)\\
&=\frac{1}{(\ii\hbar)^2}
[\hat{Q}_\text{S},\hat{S}_\text{S}] \ 
\text{Tr}_\text{A}\Big([\hat{R}_\text{A},\hat{T}_\text{A}]\rho_\text{A}\Big)
\end{align}
as claimed.

Heuristically, in a sum over histories sense, this process involves the system and ancilla interacting with each other twice via different terms in the full interaction Hamiltonian.

The general expression \eqref{SumProduct} is a direct generalization of this case.

\section{EM Dipole Cancellation}\label{EDiMDi}
In this appendix, we show that the combination of any electric multipolar coupling with the magnetic  dipole couplings cannot purify at leading order in dechoerence effect.

We begin with the simplest combination of electric and magnetic couplings
\begin{align}
H_\text{SA}
&=q \, \hat{x}^j\hat{E}_j
+\frac{q}{2m}\{\hat{L}^k,\hat{B}_k\}\\
&=\int \d\bm{x} \ \hat{d}^j(\bm{x})\otimes\hat{E}_j(\bm{x})
+\hat{\mu}^k(\bm{x})\otimes\hat{B}_k(\bm{x}).
\end{align}
The cross terms within each of the electric and magnetic couplings will vanish and only the cross terms between them will survive. Computing the effect of $\mathcal{L}_1$ on the maximally mixed states \eqref{SumProduct} yields
\be
\mathcal{L}_1[I]
\nonumber
=\Big(\frac{-\ii}{\hbar}\Big)^2\!\!\!\int\!\!\! \d\bm{x}\!\!\int\!\!\! d\bm{x'} \, 
\big\langle [\hat{E}_j(\bm{x}),\hat{B}_k(\bm{x'})]\big\rangle \, 
[\hat{d}^j(\bm{x}),\hat{\mu}^k(\bm{x'})]
\ee
This integrand is non-zero but, as as we will show the integral vanishes. 

Recall that the electric and magnetic fields have the commutator
\bel{EBCom}
[\hat{E}_i(\bm{x}),\hat{B}_j(\bm{x'})]
=\frac{-\ii\hbar}{\epsilon_0} \tensor{\varepsilon}{_i_j^ n}\nabla_n\delta(\bm{x}-\bm{x}') \ 
\openone.
\ee
where $\nabla_n=\partial/\partial x^n$ acts on the $\bm{x}$ vector.

From this we can see why this interaction cannot purify at leading order in rapid repeated interactions. Integrating by parts to move the $\nabla_n$ from the delta function onto $\bm{\hat{d}}(\bm{x})$ has the effect of transforming $\bm{\hat{d}}^i(\bm{x})\to\bm{\hat{\mu}}_j(\bm{x})$ upon using $-\ii\hbar\tensor{\varepsilon}{_i_j^ n}$.  This then leads to a vanishing commutator.
Using \eqref{EBCom} we have,
\begin{align}
\mathcal{L}_1[I]
=&\nonumber
\frac{1}{\hbar^2}\!\!
\int\!\!\!\d\bm{x}\!\!\int\!\!\!\d\bm{x'} \ 
\Big\langle \frac{-\ii\hbar}{\epsilon_0} \tensor{\varepsilon}{_i_j^ n}\nabla_n\delta(\bm{x}-\bm{x'})\boldsymbol{\hat{1}}\Big\rangle
[\hat{d}^i(\bm{x}),\hat{\mu}^j(\bm{x'})]\\
=&\nonumber
\frac{-1}{\hbar^2}\!\!
\int\!\!\!\d\bm{x}\!\!\int\!\!\!\d\bm{x'} \ 
\Big\langle \frac{-\ii\hbar}{\epsilon_0} \tensor{\varepsilon}{_i_j^ n}\delta(\bm{x}-\bm{x'})\boldsymbol{\hat{1}}\Big\rangle
[\nabla_n\hat{d}^i(\bm{x}),\hat{\mu}^j(\bm{x'})]\\
=&\nonumber
\frac{-1}{\hbar^2\epsilon_0}\!\!
\int\!\!\!\d\bm{x}\!\!\int\!\!\!\d\bm{x}' \ 
\delta(\bm{x}-\bm{x'}) \big\langle
\boldsymbol{\hat{1}}\big\rangle
[-\ii\hbar \, \tensor{\varepsilon}{_i_j^ n}\nabla_n\hat{d}^i(\bm{x}),\hat{\mu}^j(\bm{x'})]\\
=&\label{EdipoleBdipoleInt}
\frac{-1}{\hbar^2\epsilon_0}\!\!
\int\!\!\!\d\bm{x} \, 
[-\ii\hbar \, \tensor{\varepsilon}{_i_j^ n}\nabla_n\hat{d}^i(\bm{x}),\hat{\mu}^j(\bm{x})].
\end{align}
and (as we shall demonstrate), since
\bel{dtomu}
-\ii\hbar \, \tensor{\varepsilon}{_i_j^ n}\nabla_n\hat{d}^i(\bm{x})
=-2\hat{\mu}_j(\bm{x})
\ee
 the commutator thus vanishes. This is not unexpected since $\bm{\hat{d}}\sim\bm{\hat{x}}$ and $\bm{\hat{\mu}}\sim\bm{\hat{L}}$. 

In order to show \eqref{dtomu} we must first note that the Levi-Cevita symbol, $\tensor{\varepsilon}{_i_j^ n}$, forces $i\neq n$ such that $x^i$ and $\nabla_n$ commute. Secondly we must recognize that 
\begin{align}
-\ii\hbar\nabla_n\big(\ket{\bm{x}}\!\bra{\bm{x}}\big)
=\big\{\hat{p}_n,\ket{\bm{x}}\!\bra{\bm{x}}\big\}
\end{align}
which can be seen by computing
\begin{align}
&\bra{\psi}-\ii\hbar\nabla_n\big(\ket{\bm{x}}\!\bra{\bm{x}}\big)\ket{\phi}\\
&\nonumber
=-\ii\hbar\nabla_n\big(\braket{\psi|\bm{x}}\braket{\bm{x}|\phi}\big)\\
&\nonumber
=-\ii\hbar\nabla_n\big(\psi^*(\bm{x})\phi(\bm{x})\big)\\
&\nonumber
=\big(-\ii\hbar\nabla_n\psi^*(\bm{x})\big)\phi(\bm{x})
+\psi^*(\bm{x})\big(-\ii\hbar\nabla_n\phi(\bm{x})\big)\\
&\nonumber
=\bra{\psi}\hat{p}_n\ket{\bm{x}}\braket{\bm{x}|\phi}
+\braket{\psi|\bm{x}}\bra{\bm{x}}\hat{p}_n\ket{\psi}\\
&\nonumber
=\bra{\psi}\big\{\hat{p}_n,\ket{\bm{x}}\!\bra{\bm{x}}\big\}\ket{\phi}.
\end{align}
Using these two results we can straightforwardly compute
\begin{align}
-\ii\hbar \, \tensor{\varepsilon}{_i_j^ n}\nabla_n\hat{d}^i(\bm{x})
&=-\ii\hbar \, \tensor{\varepsilon}{_i_j^ n}\nabla_n\big(q \, x^i\ket{\bm{x}}\!\bra{\bm{x}}\big)\\
&\nonumber
=-\ii\hbar \, \tensor{\varepsilon}{_i_j^ n}
q \,\hat{x}^i\nabla_n\big(\ket{\bm{x}}\!\bra{\bm{x}}\big)\\
&\nonumber
=q \,\tensor{\varepsilon}{_i_j^ n}
\hat{x}^i\{\hat{p}_n,\ket{\bm{x}}\!\bra{\bm{x}}\}\\
&\nonumber
=q \,\{\tensor{\varepsilon}{_i_j^ n}\hat{x}^i\hat{p}_n,\ket{\bm{x}}\!\bra{\bm{x}}\}\\
&\nonumber
=-q \,\{\hat{L}_j,\ket{\bm{x}}\!\bra{\bm{x}}\}\\
&\nonumber
=-2 \, m \, \hat{\mu}_j(\bm{x})
\end{align}
Thus the commutator in \eqref{EdipoleBdipoleInt} vanishes.

In fact, taking a combination of both the electric quadrupole and magnetic couplings as,
\begin{align}
H_\text{SA}
&=q \, \hat{x}^i\hat{x}^j\nabla_i\hat{E}_j
+\frac{q}{2m}\{\hat{L}^k,\hat{B}_k\}\\
&=\int \d\bm{x} \ \hat{Q}^i{}^j(\bm{x})\otimes\nabla_i\hat{E}_j(\bm{x})
+\hat{\mu}^k(\bm{x})\otimes\hat{B}_k(\bm{x}).
\end{align}
we find a similar cancellation such that will yield no purification at leading order. 
Any higher order electric couplings will exibit the same cancellation as will the combination of several electric multipolar moments along with the magnetic dipole moment. Thus if there are any light-atom interactions capable of purifying at leading order they must involve quadrupolar or higher magnetic couplings.

\bibliography{references}
\end{document}